\documentclass{PoS}

\title{News on spectra from the NA61/SHINE experiment}

\ShortTitle{Spectra from NA61/SHINE}

\author{\speaker{Magdalena Kuich}\thanks{This work was partially supported by the National Science Centre, Poland (grant 2015/18/M/ST2/00125).} \hspace{0.1cm} for the  NA61/SHINE Collaboration \\
        University of Warsaw \\
        E-mail: \email{magdalena.kuich@cern.ch}}

\abstract{NA61/SHINE is a fixed target experiment at the CERN Super-Proton-Synchrotron. The main goals of the experiment are to discover the critical point of strongly interacting matter and study the properties of the onset of deconfinement. In order to reach these goals, a study of hadron production properties is performed in nucleus-nucleus, proton-proton and proton-nucleus interactions as a function of collision energy and size of the colliding nuclei.
In this talk, recent results on particle production in p+p interactions, as well as Be+Be and Ar+Sc collisions in the SPS energy range are reviewed. Transverse momentum, transverse mass and rapidity spectra obtained with various analysis methods are presented. Surprises in studies of signatures of onset of deconfinement are discussed. The results are compared with available world data.}

\FullConference{The European Physical Society Conference on High Energy Physics\\
          5-12 July\\
          Venice, Italy} 

\begin{document}

\section{Introduction}
The NA61/SHINE experiment performs a unique two-dimensional scan of the phase diagram of strongly interacting matter. The main goals are the study of the properties of the onset of deconfinement by measurements of hadron production and the search for the critical point of strongly interacting matter by measuring event-by-event fluctuations. Measurements are performed at CERN SPS beam momenta (13$A$-150/158$A$ GeV/$c$) for various system sizes (p+p, p+Pb, Be+Be, Ar+Sc, Xe+La and Pb+Pb).
The programme is motivated by the discovery of the  onset of deconfinement in Pb+Pb collisions at 30$A$ GeV/c by the NA49 experiment \cite{1, 2}. 

\section{NA61/SHINE experiment}
NA61/SHINE is a fixed target experiment at the CERN SPS \cite{3}. The detection system is based on eight Time Projection Chambers (TPC) providing acceptance in the full forward center-of-mass hemisphere, down to $p_\textup{T}=0$. TPCs allow for track, momentum and charge reconstruction as well as mean energy loss (dE/dx) measurement per unit path length. Time of Flight walls provide additional particle identification measuring particles mass in smaller acceptance regions. A zero-degree calorimeter, Projectile Spectator Detector, allows to select central collisions based on the measurement of the forward energy mostly contributed by projectile spectator nucleons and fragments.

\section{Recent results on study of the onset of deconfinement}
\subsection{Negatively charged pion spectra}
\begin{figure}
\centering
\begin{minipage}{.4\textwidth}
\includegraphics[width=.9\textwidth]{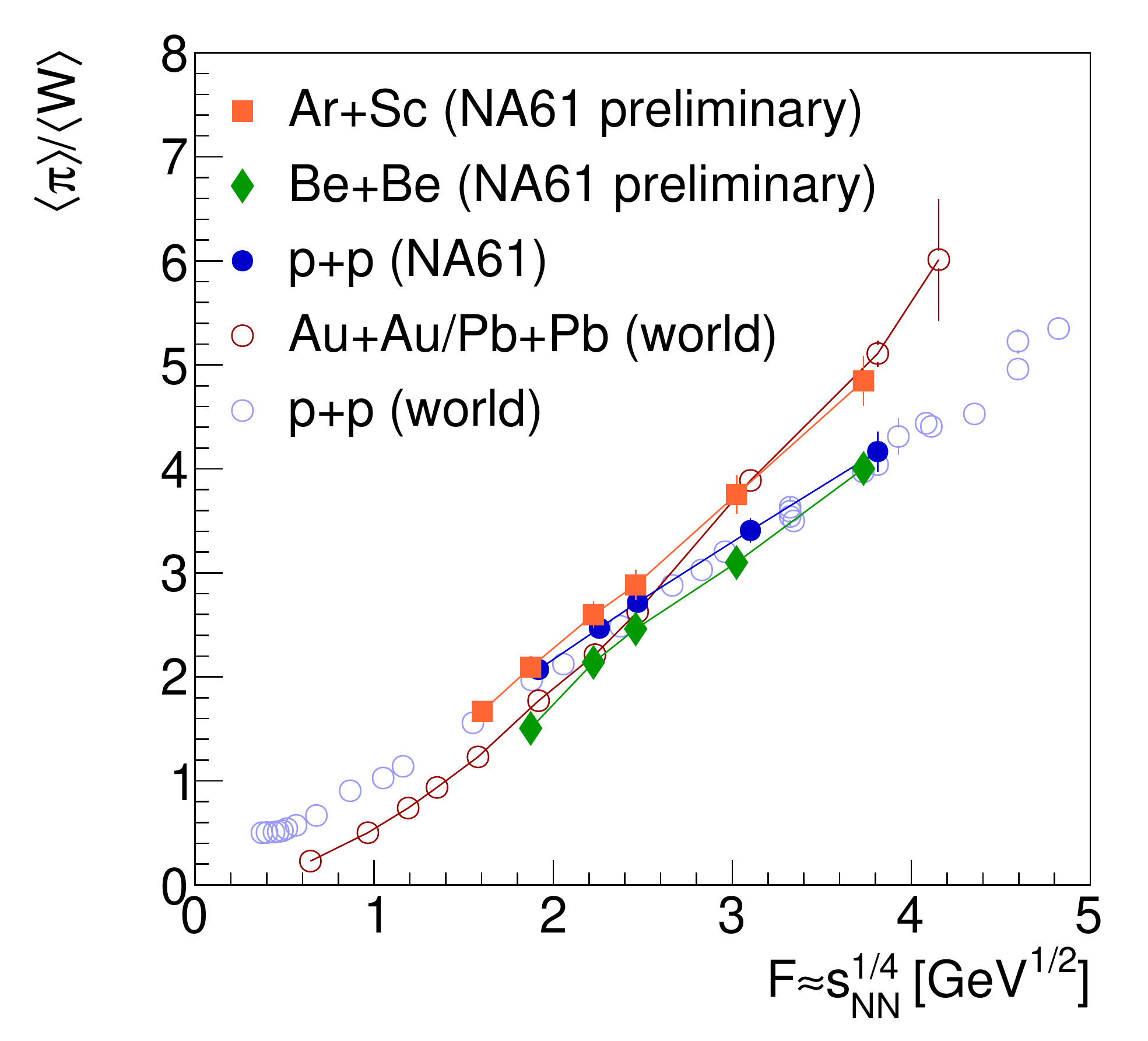}
\end{minipage}
\begin{minipage}{.4\textwidth}
\caption{Energy dependence of the total pion multiplicity $\pi$ normalised to the mean number of wounded nucleons in the reaction.}
\label{fig:kink}
\end{minipage}
\end{figure}

The negatively charged pion spectra in p+p \cite{4}, central Be+Be \cite{5, 6} and central Ar+Sc collisions \cite{7, 8, 9} were derived in large acceptance from unidentified negatively charged hadron spectra using the so-called $h^-$ method. The method is based on the fact that the majority of negatively charged particles produced in p+p and ion+ion collisions are pions. The small non-pion contribution is subtracted from the spectra based on estimates obtained with the EPOS 1.99 model. 
\begin{figure}
\begin{minipage}{.72\textwidth}
\includegraphics[width=.5\textwidth]{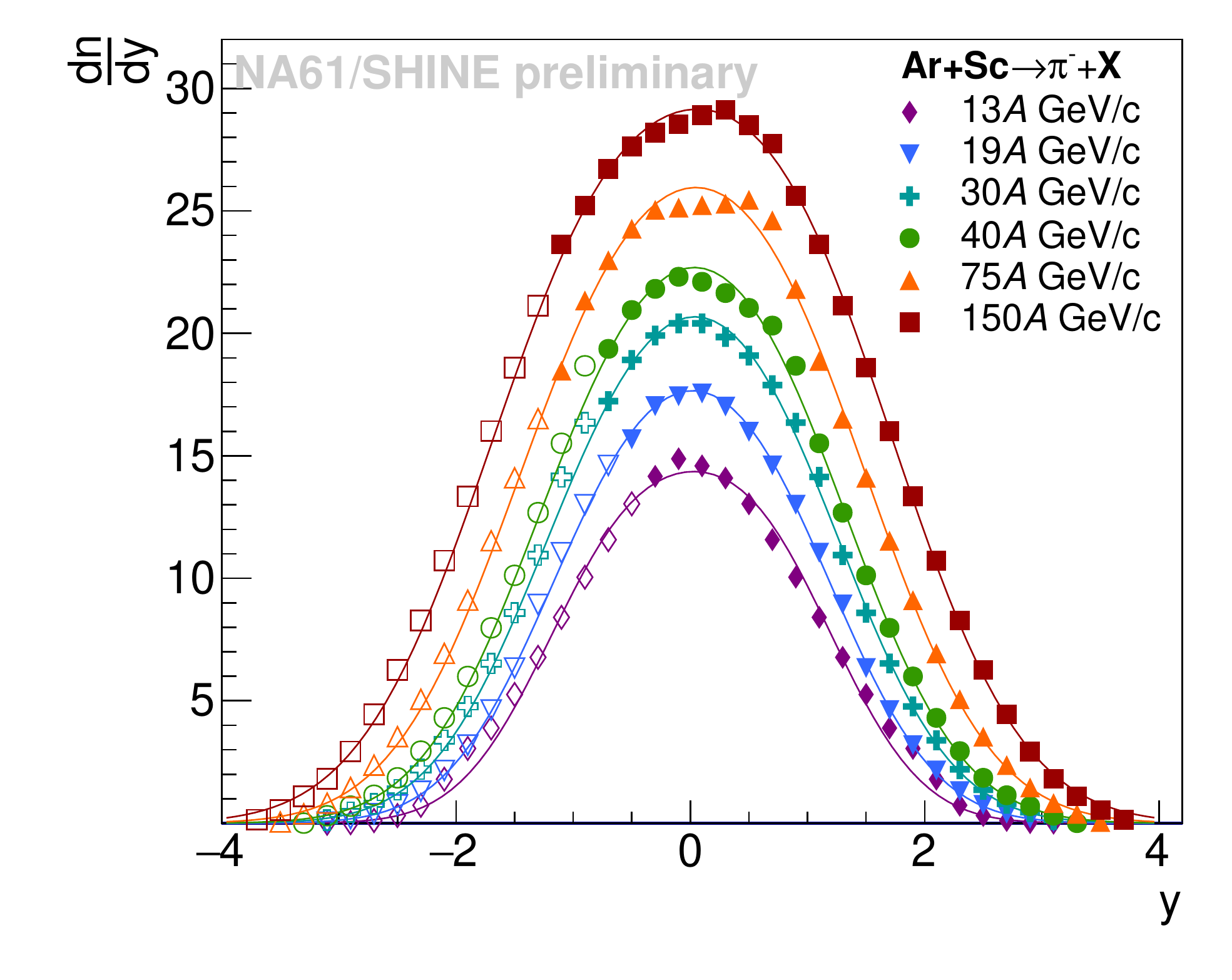}\includegraphics[width=.41\textwidth]{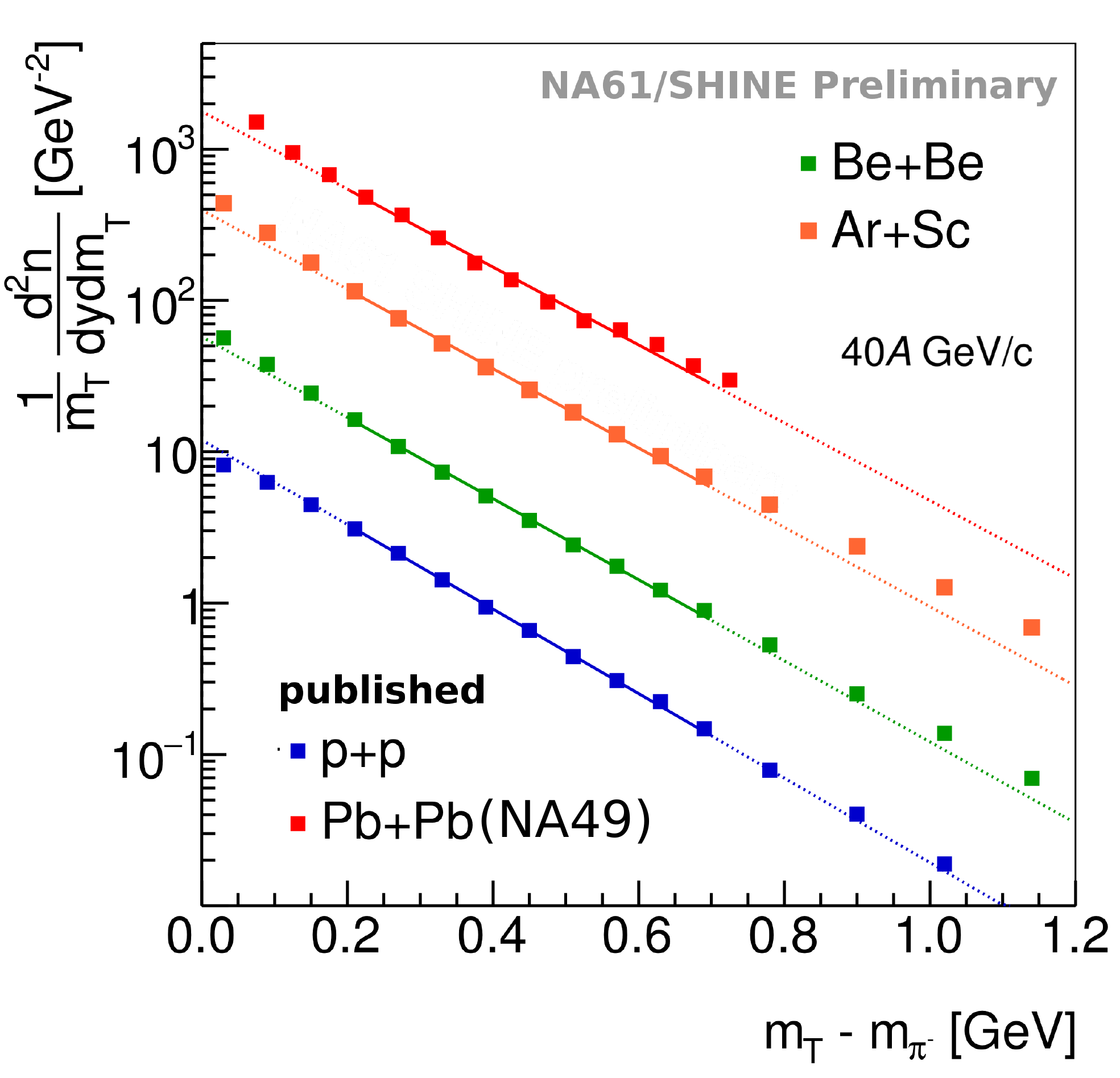}
\end{minipage}
\begin{minipage}{.27\textwidth}
\caption{Negatively charged pion spectra. Left:  Rapidity spectra in Ar+Sc collisions at six beam momenta. Right: Transverse mass spectra at mid-rapidity at 40A GeV/c for various collision systems. An exponential function (line) was fitted in $0.2<m_\textup{T} - m_
\pi<0.7$GeV/$c$. }
\label{fig:hminusy}
\end{minipage}
\end{figure}
Figure \ref{fig:hminusy} (left) shows rapidity spectra in Ar+Sc collisions at six beam momenta. Rapidity spectra are approximately gaussian, independently of the collision energy. The large detector acceptance allows obtaining total yields of pions from the measured data and minor extrapolations (for details see Ref. \cite{8}). Transverse mass spectra at mid-rapidity at 40$A$ GeV/$c$ in p+p, Be+Be and Ar+Sc interactions are shown in Fig. \ref{fig:hminusy} (right) along with Pb+Pb (NA49) \cite{2} results for comparison. The $m_\textup{T}$ spectra are exponential in p+p interactions, but deviate from this shape for heavier systems. Nevertheless, exponential functions were fitted in bins of rapidity to all systems in the $m_\textup{T} - m_\pi$ range 0.2-0.7 GeV in order to extrapolate to the unmeasured region.
Mean multiplicities of all pions ($\langle \pi \rangle$) normalised to the average number of wounded nucleons ($\langle W \rangle$) are shown in Fig. \ref{fig:kink}. Results are compared with results from other experiments \cite{10, 11, 12}. At higher SPS energies the rate of increase with collision energy is larger for the heavy systems (Pb+Pb, Ar+Sc) than for the light ones (p+p, Be+Be). The Statistical Model of the Early Stage (SMES) predicts such an increase in the quark-gluon plasma due to the larger number of degrees of freedom \cite{13} compared to a hadron gas.

\subsection{Charged hadron spectra}
\begin{figure}
\includegraphics[width=.38\textwidth,trim={0.5cm 0 0 0},clip]{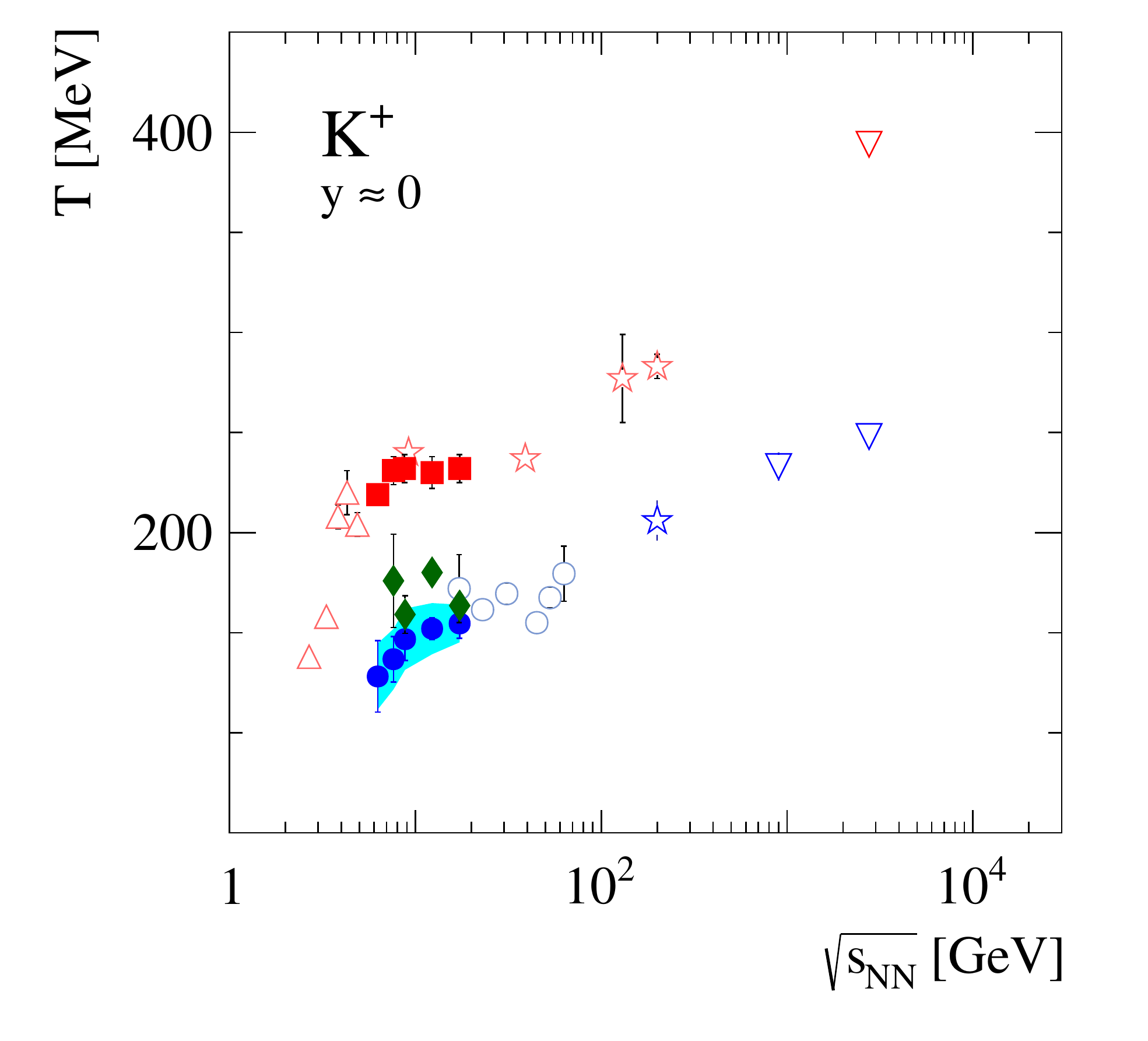}
\includegraphics[width=.38\textwidth,trim={0.5cm 0 0 0},clip]{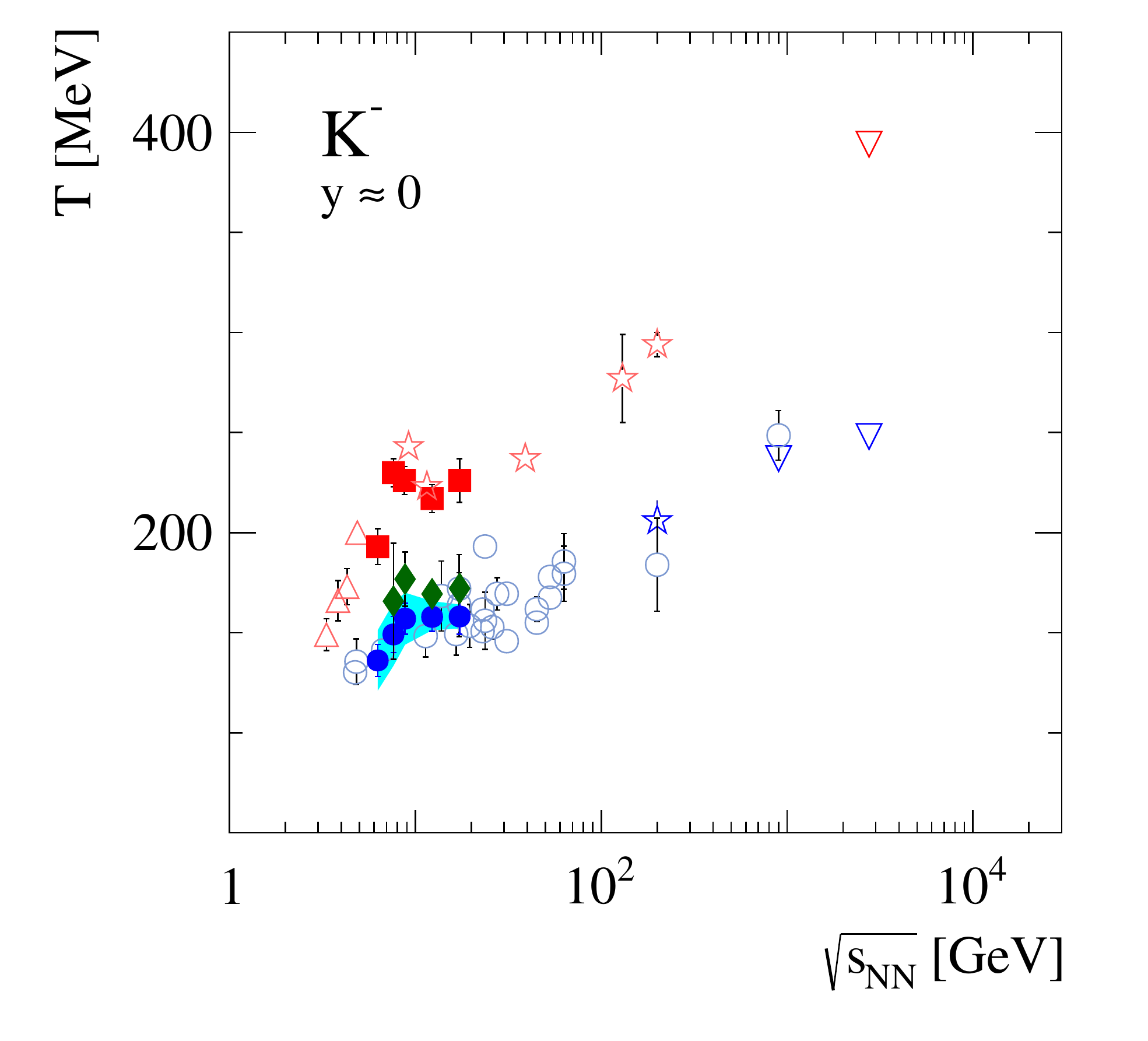}
\includegraphics[width=.2\textwidth,trim={1.7cm 0 8.1cm 0},clip]{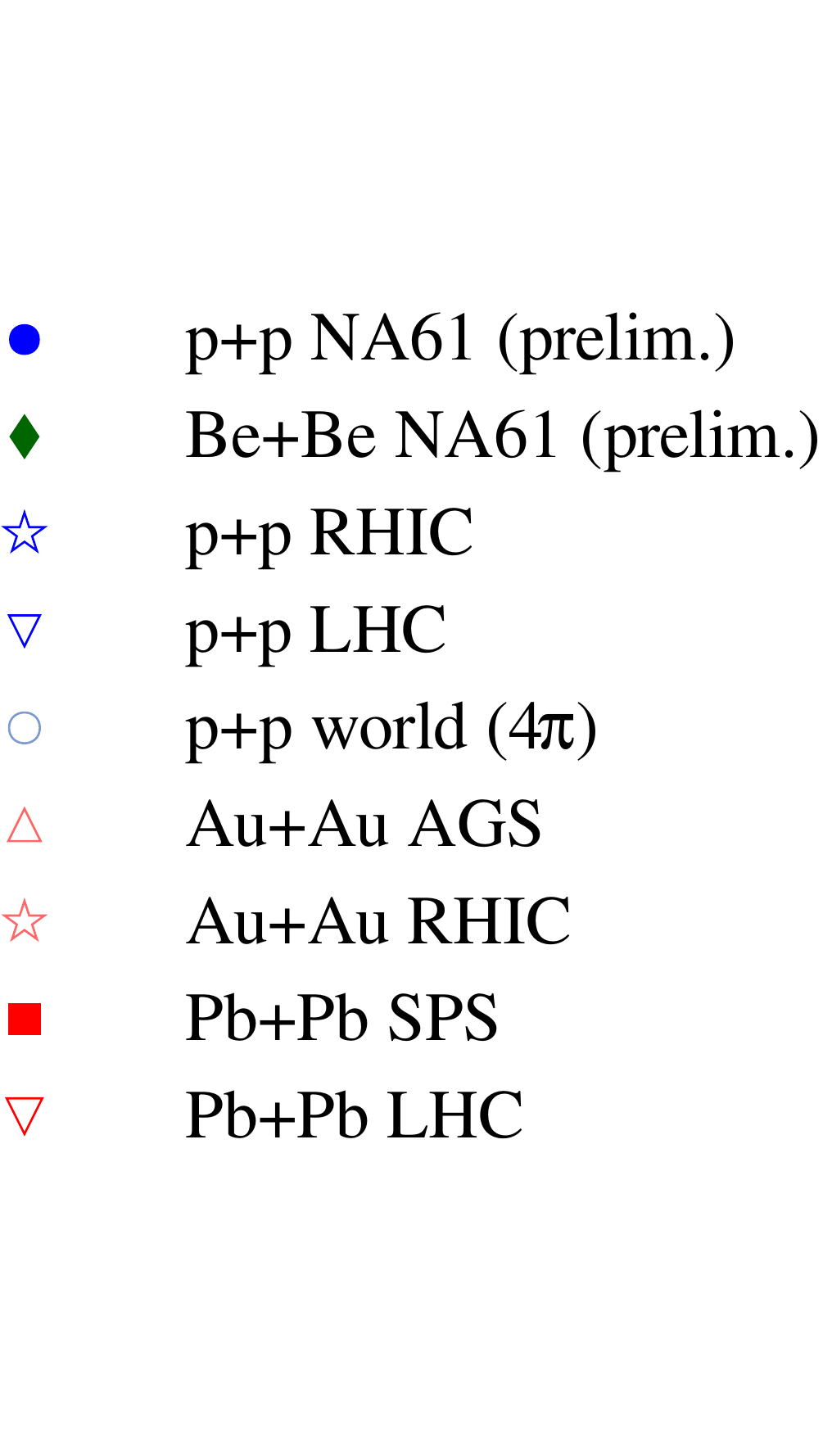}
\caption{Energy dependence of the inverse slope parameter of $m_\textup{T}$ spectra at mid-rapidity of positively (left) and negatively (right) charged kaons.}
\label{fig:step}
\end{figure}

\begin{figure}
\includegraphics[width=.38\textwidth,trim={0.5cm 0 0 0},clip]{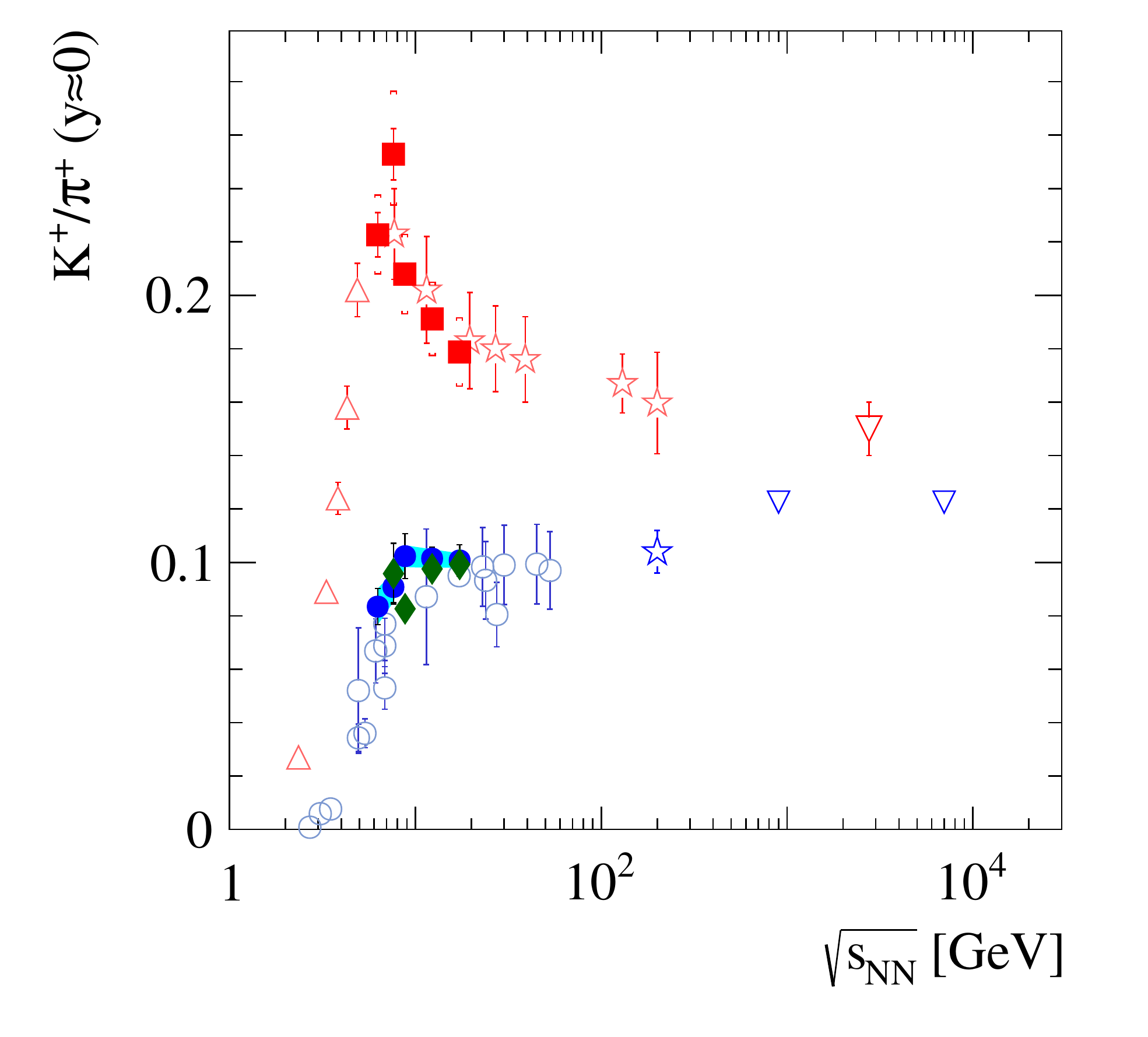}
\includegraphics[width=.38\textwidth,trim={0.5cm 0 0 0},clip]{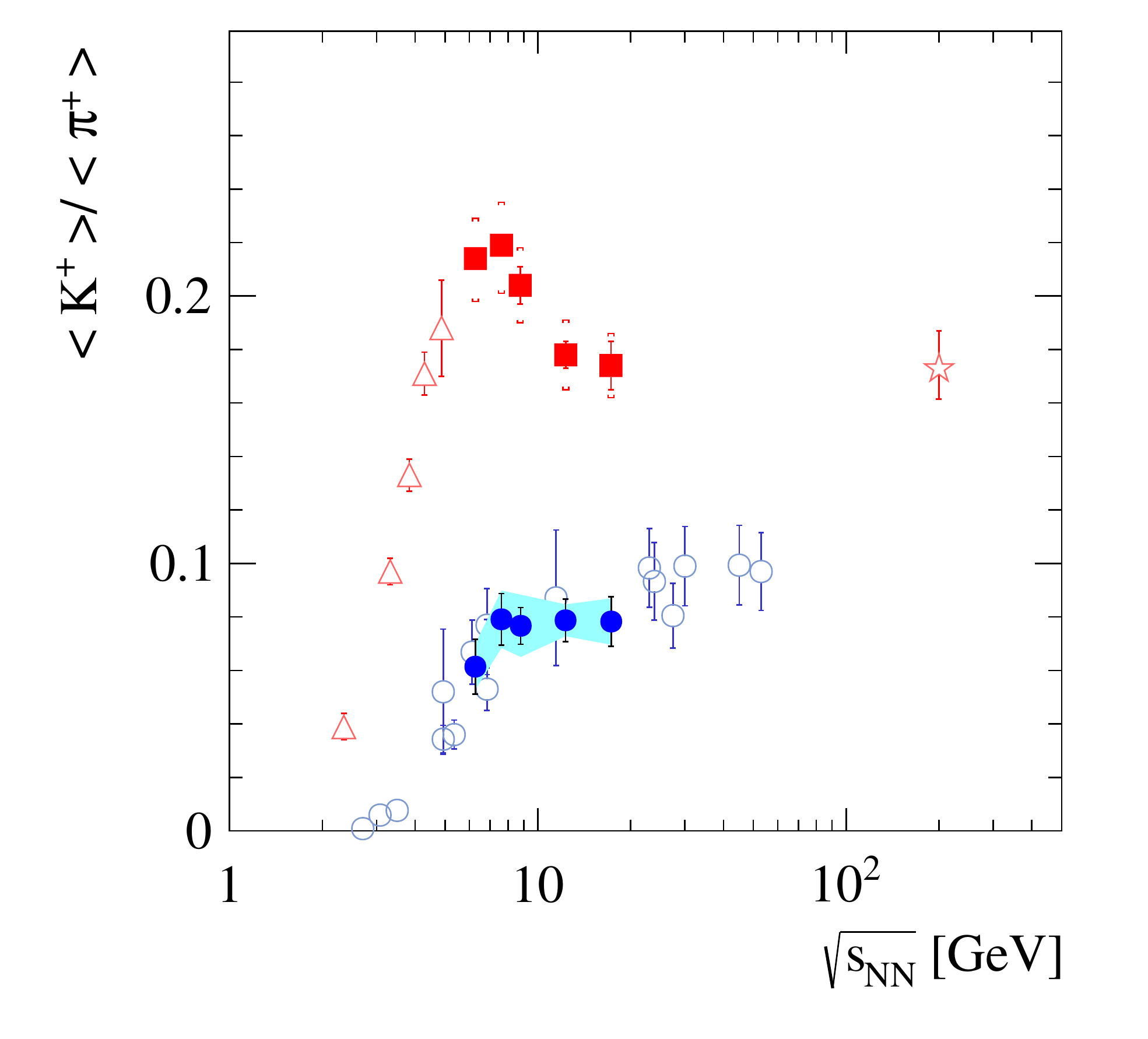}
\includegraphics[width=.2\textwidth,trim={1.7cm 0 8.1cm 0},clip]{./pictures/legenda}
\caption{Energy dependence of the positively charged kaon to pion yield ratio at mid-rapidity (left) and in $4\pi$ acceptance (right).}
\label{fig:horn}
\end{figure}

Charged hadrons ($\pi^\pm$, p, $\overline{\textup{p}}$ and K$^\pm$) in p+p interactions and K$^\pm$ near mid-rapidity from central Be+Be collisions were identified based on the measurements of the energy loss in the TPCs (dE/dx) and time of flight in the ToF detectors. Figure~\ref{fig:step} presents the energy dependence of the inverse slope parameter of $m_\textup{T}$ spectra of charged kaons. Figure~\ref{fig:horn} shows the multiplicity ratio of charged kaons to pions at mid-rapidity. In both figures NA61/SHINE results from p+p interactions \cite{14} and Be+Be collisions are compared with results from central Pb+Pb collisions from NA49 \cite{1, 2} and other experiments \cite{15, 16, 17, 18, 19, 20, 21, 22}. 

A plateau visible in the energy dependence of the inverse slope parameter in Fig.~\ref{fig:step} and peaks seen in Fig.~~\ref{fig:horn} for Pb+Pb and Au+Au collisions in the SPS energy range were predicted by the SMES model as signatures of the onset of deconfinement. In the SPS energy range the NA61/SHINE results on p+p interactions exhibit a qualitatively similar energy dependence (step) for the inverse splope parameter and a step instead of a peak in the kaon to pion ratio. Thus some properties of hadron production previously attributed to the onset of deconfinement in heavy ion collisions are present also in p+p interactions. Surprisingly, while the inverse slope parameter in Be+Be collisions lies slightly above the one
in p+p interactions, the values of the charged kaon to pion ratio are very close in Be+Be and p+p. 
\begin{figure}
\centering
\begin{minipage}{.77\textwidth}
\includegraphics[width=.45\textwidth]{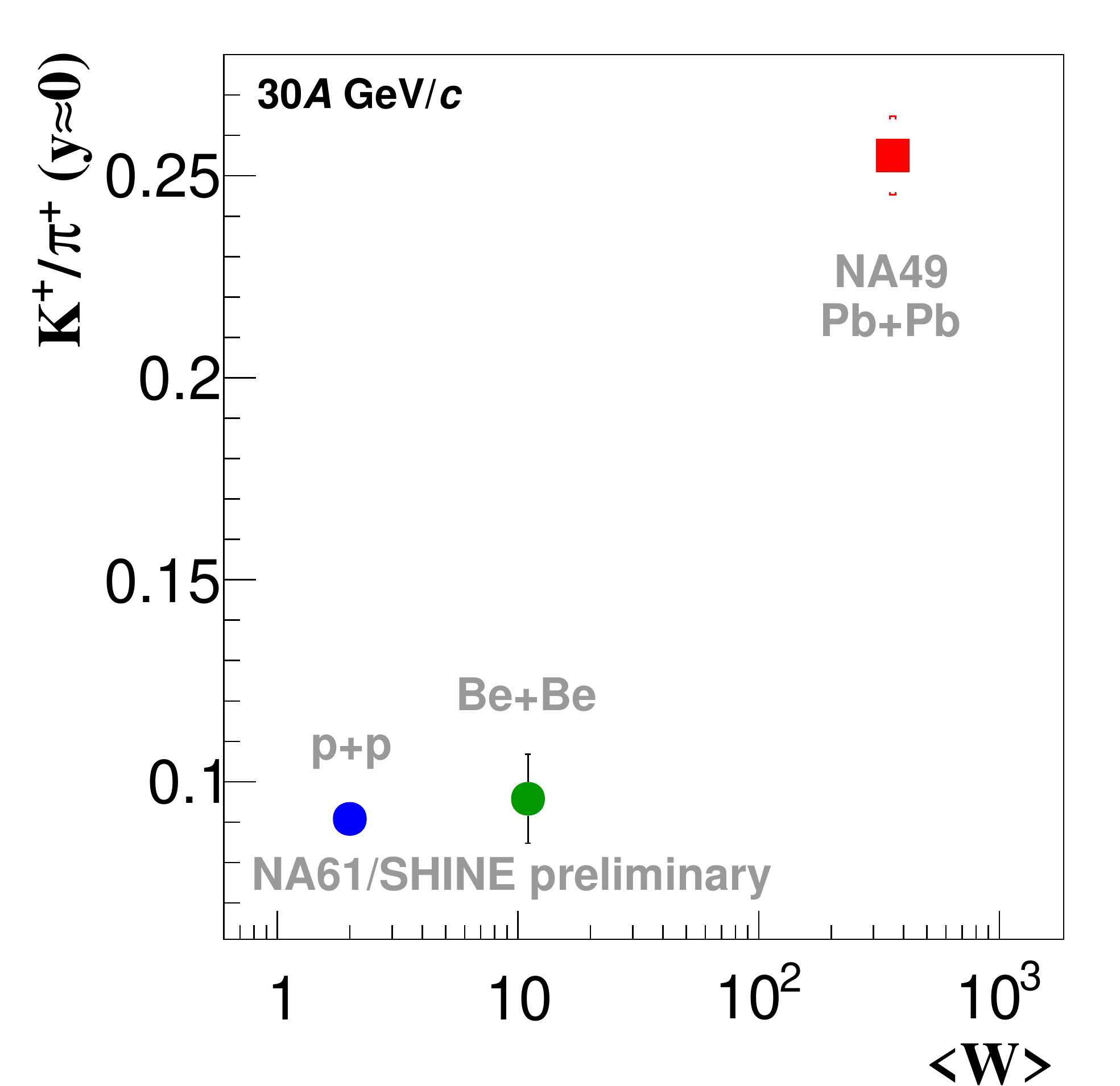}
\includegraphics[width=.45\textwidth]{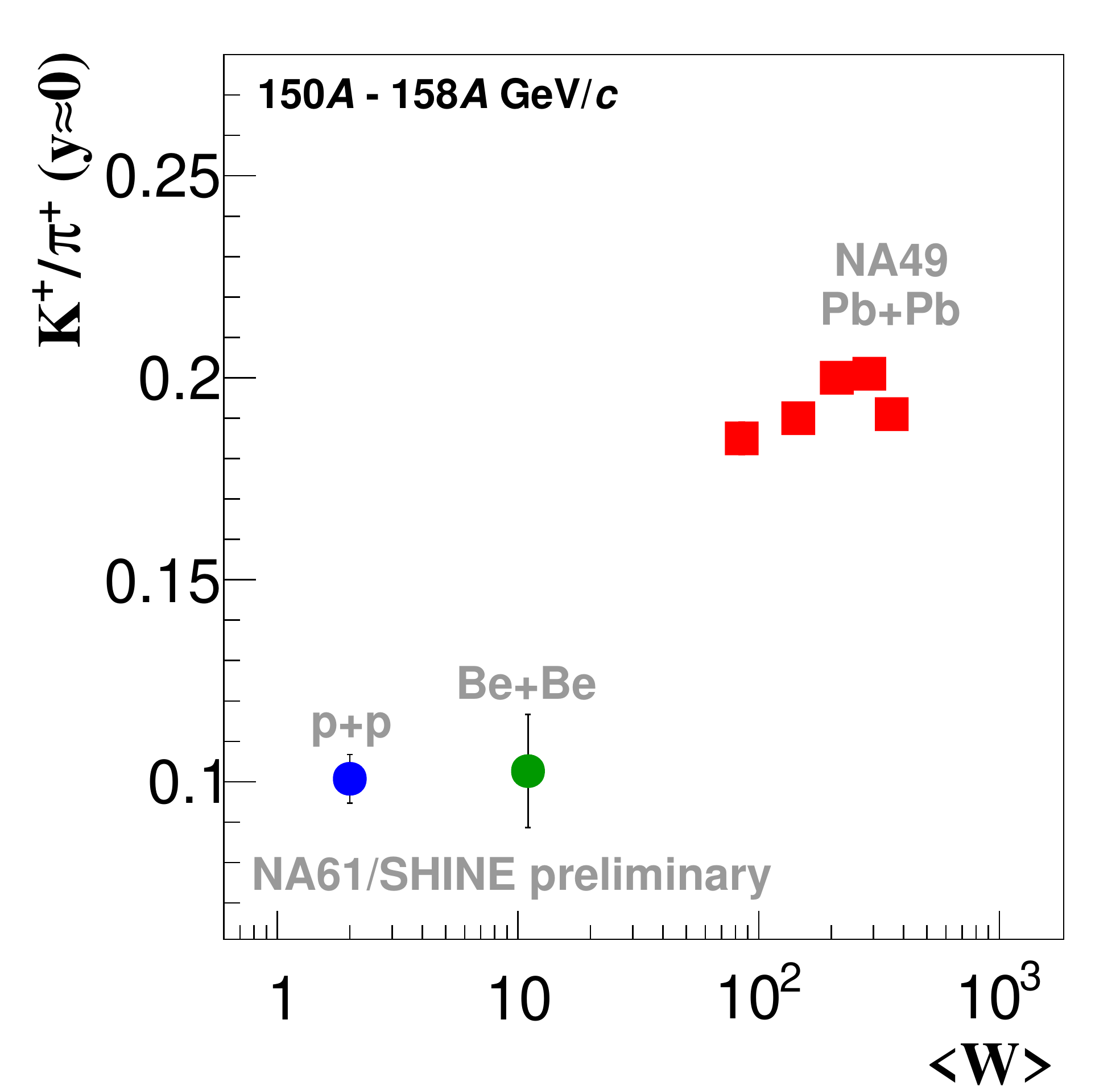}
\end{minipage}
\begin{minipage}{.22\textwidth}
\caption{System size dependence of the positively charged kaon to pion yield ratio at mid-rapidity at 30$A$ (left) and 150$A$ (right) GeV/$c$ beam momentum.}
\label{fig:systemdep}
\end{minipage}
\end{figure}
Figure \ref{fig:systemdep} presents the system size dependence of the positively charged kaon to pion yield ratio at mid-rapidity at 30$A$ (left) and 150$A$ (right) GeV/$c$ beam momentum. A clear jump between light systems (p+p and Be+Be) and a heavy system (Pb+Pb) suggests a threshold behaviour of quark-gluon plasma production in system size.

\subsection{$\Lambda$ hyperon spectra}
\begin{figure}
\centering
\begin{minipage}{.78\textwidth}
\includegraphics[width=\textwidth, trim={0 0 0 0.5cm}, clip]{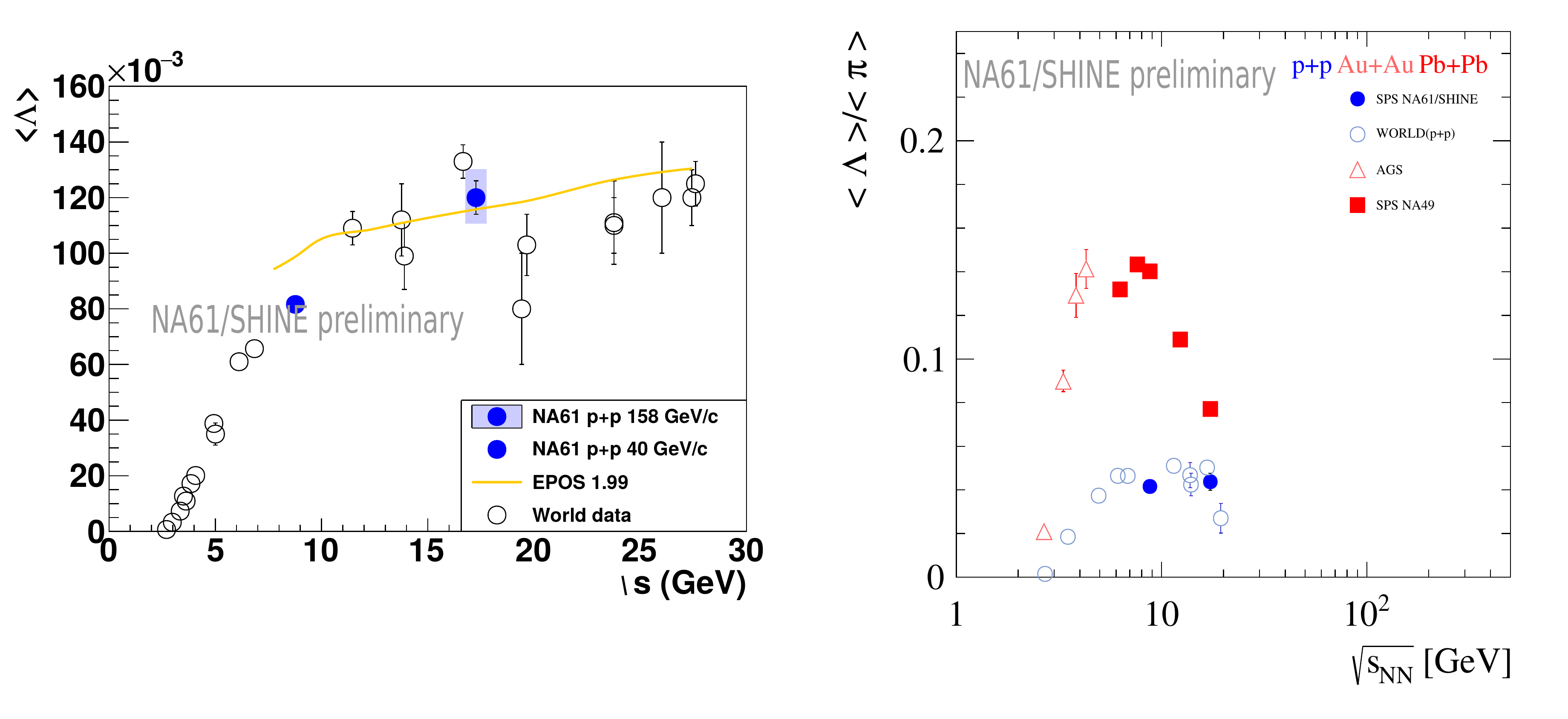}
\end{minipage}
\begin{minipage}{.21\textwidth}
\caption{Energy dependence of $\Lambda$ yield (left) and $\Lambda$ to $\pi$ ratio (right) at 40 and 150 GeV/c compared to p+p and ion-ion world data.}
\label{fig:lambda1}
\end{minipage}
\end{figure}
NA61/SHINE measured two-dimensional (rapidity versus transverse momentum) sectra of $\Lambda$ hyperons in p+p interactions at 40 \cite{23} and 158 GeV/c \cite{24}. Figure \ref{fig:lambda1} (left) presents the energy dependence of the total $\Lambda$ yield compared with world results from p+p interactions and EPOS 1.99 model predictions. The right panel of Fig. \ref{fig:lambda1} shows the energy dependence of the ratio of total $\Lambda$ to $\pi$ multiplicity, compared with other results for p+p and heavy ion collisions. The ratio shows a maximum in the SPS energy range for heavy ions, which is similar to the maximum visible in the ratio of total K to $\pi$ multiplicities (see right panel of Fig. \ref{fig:horn}). In contrast to Pb+Pb the $\Lambda / \pi$ ratio in p+p interactions reaches a plateau at the SPS energies.

\subsection{$\phi$ meson spectra}
\begin{figure}
\begin{minipage}{.8\textwidth}
\includegraphics[width=.49\textwidth]{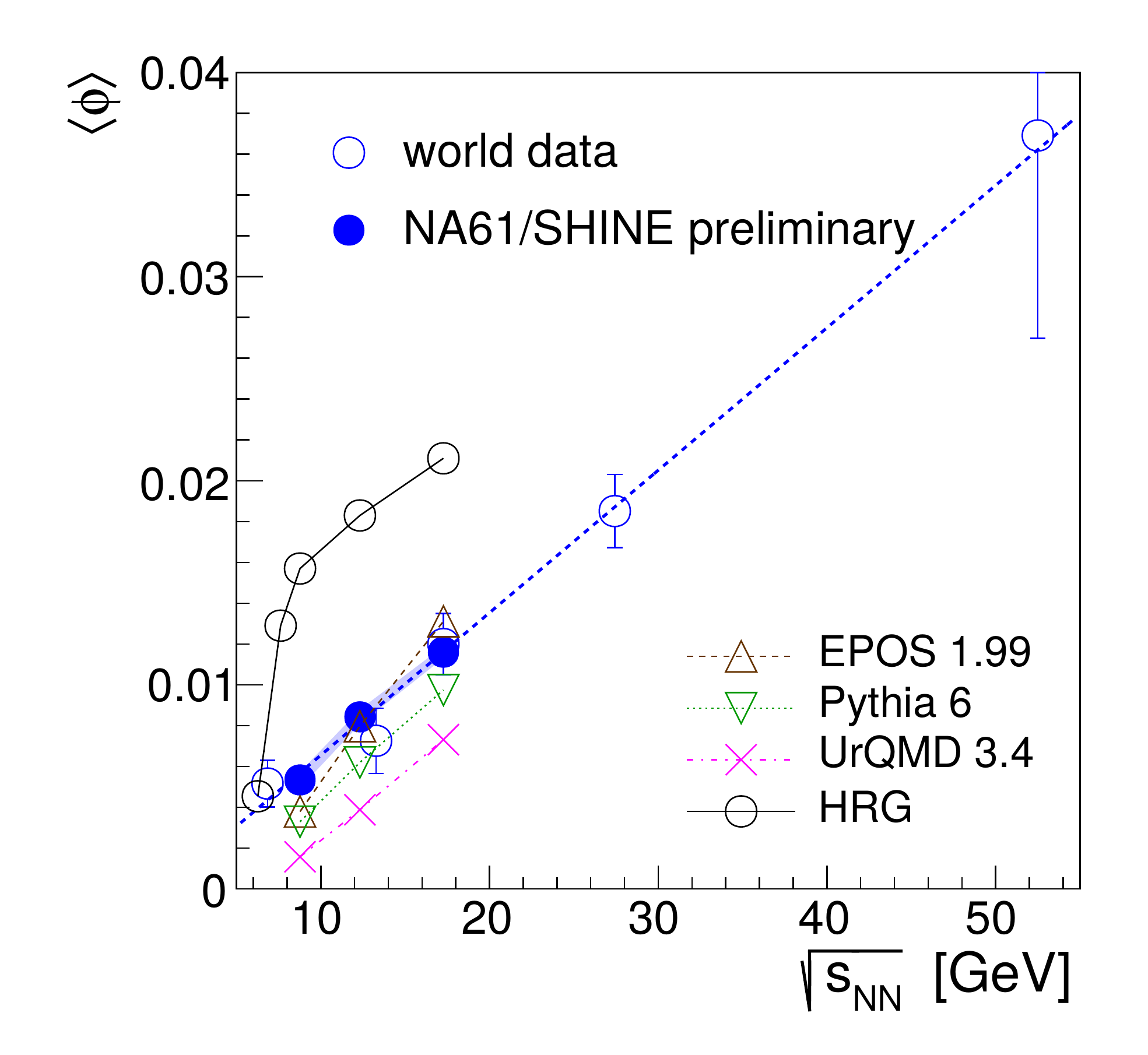}\includegraphics[width=.49\textwidth]{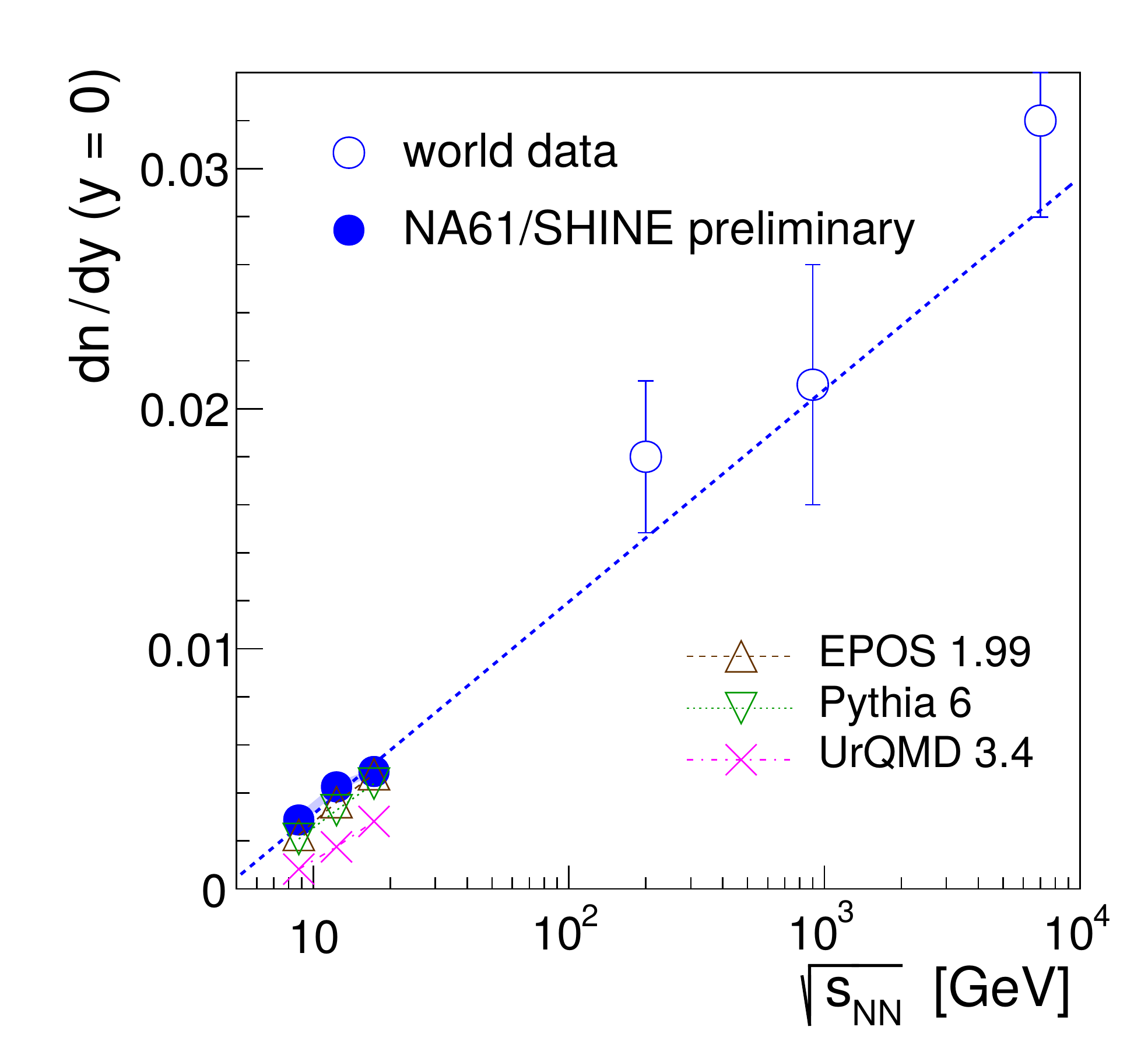}
\end{minipage}
\begin{minipage}{.19\textwidth}
\caption{Energy dependence of $\phi$ production in p+p interactions: total yields (left) and mid-rapidity yield (right).}
\label{fig:phitotal}
\end{minipage}
\end{figure}

NA61/SHINE also performs measurements of yields of heavier mesons. Figure \ref{fig:phitotal} presents the energy dependence of $\phi$ meson production. The mean $\phi$ multiplicity is plotted in Fig. \ref{fig:phitotal} (left) and compared to various model predictions. The best agreement with the data is given by the EPOS 1.99 model, however, the predicted energy dependence seems to be too steep. Figure \ref{fig:phitotal} (right) shows the energy dependence of mid-rapidity yields which are well described by the EPOS 1.99 and Pythia 6 models.

\section{Summary}
This contribution focused on recent results from the NA61/SHINE strong interactions program aiming to study the onset of deconfinement. Results on spectra were presented, in particular the new charged kaon spectra in Be+Be collisions at 30$A$--150$A$ GeV/c. The inverse slope parameter of the transverse mass distribution of charged kaons is higher in Be+Be than in p+p interactions, while the charged kaon to pion multiplicity ratio is at a similar level much below that seen in Pb+Pb collisions. This raises questions about the mechanism of onset of deconfinement as a function of system size. Moreover, new $\phi$ production measurements in p+p interactions at 40, 80 and 158 GeV/$c$ beam momenta were reported. The two-dimensional scan of particle production at SPS energies with size of the collision system will be completed in 2018. As an extension of this program NA61/SHINE
plans to measure precisely open charm and multi-strange hyperon production in 2021--2024.

\end{document}